\documentclass[preprint]{aastex}
\usepackage{emulateapj5}
\usepackage{apjfonts}

\input psfig.tex
\usepackage{natbib}
\def\amin{$^\prime$}

\def\apjs{{ApJS}}
\def\asec{$^{\prime\prime}$}

\def\ha{H$\alpha$}
\def\hb{H$\beta$}
\def\hst{{\it HST}}
\def\galex{{\it GALEX}}
\def\kms{km s$^{-1}$}
\def\lamb{$\lambda$}
\def\lax{{$\mathrel{\hbox{\rlap{\hbox{\lower4pt\hbox{$\sim$}}}\hbox{$<$}}}$}}
\def\gax{{$\mathrel{\hbox{\rlap{\hbox{\lower4pt\hbox{$\sim$}}}\hbox{$>$}}}$}}
\def\simlt{\lower.5ex\hbox{$\; \buildrel < \over \sim \;$}}
\def\simgt{\lower.5ex\hbox{$\; \buildrel > \over \sim \;$}}
\def\lum{erg s$^{-1}$}
\def\mbh{{$M_{\rm BH}$}}

\def\percm2{cm$^{-2}$}

\def\solmass{$M_\odot$}
\def\oii{[\ion{O}{2}]}

\def\hi{\ion{H}{1}}

\def\oiii{[\ion{O}{3}]}

\def\nii{[\ion{N}{2}]}

\def\ser{S\'{e}rsic}

\shorttitle{ULX in NGC 5252}
\shortauthors{KIM ET AL.}

\begin{document}

\title{An Off-nucleus Non-stellar Black Hole in the Seyfert Galaxy NGC 5252}


\author{Minjin Kim\altaffilmark{1,2}, 
Luis C. Ho\altaffilmark{3,4},
Junfeng Wang\altaffilmark{5}, 
Giuseppina Fabbiano\altaffilmark{6}, 
Stefano Bianchi\altaffilmark{7},
Massimo Cappi\altaffilmark{8},
Mauro Dadina\altaffilmark{8},
Giuseppe Malaguti\altaffilmark{8},
and Chen Wang\altaffilmark{5}
}

\altaffiltext{1}{Korea Astronomy and Space Science Institute, Daejeon 305-348, 
Republic of Korea}

\altaffiltext{2}{University of Science and Technology, Daejeon 305-350, 
Republic of Korea}

\altaffiltext{3}{Kavli Institute for Astronomy and Astrophysics, 
Peking University, Beijing 100871, China} 

\altaffiltext{4}{Department of Astronomy, School of Physics, 
Peking University, Beijing 100871, China} 

\altaffiltext{5}{Department of Astronomy and Institute of Theoretical Physics 
and Astrophysics, Xiamen University, Xiamen, Fujian 361005, China}

\altaffiltext{6}{Harvard-Smithsonian Center for Astrophysics, 60 Garden 
Street, Cambridge, MA 02138, US}

\altaffiltext{7}{Dipartimento di Matematica e Fisica, Universit\`{a} degli 
Studi Roma Tre, via della Vasca Navale 84, 00146 Roma, Italy}

\altaffiltext{8}{INAF-IASF Bologna, I-40129 Bologna, Italy}

\begin{abstract}
We report the discovery of a ultraluminous X-ray source (ULX; CXO 
J133815.6+043255) in NGC 5252. This ULX is an off-nuclear point-source, which 
is 22\asec\ away from the center of NGC 5252, and has an X-ray luminosity of 
1.5 
$\times$ $10^{40}$ \lum. It is one of the rare examples of ULX, which exhibits 
clear counterparts in radio, optical, UV bands. Follow-up optical spectrum of 
the ULX shows strong emission lines.
The redshift of \oiii\ emission line coincides with the systematic velocity of 
NGC 5252, suggesting the ULX is gravitationally bound to NGC 5252. The flux of 
\oiii\ appears to be correlated with both X-ray and radio luminosity 
in the same manner as ordinary AGNs, indicating that the \oiii\ emission is 
intrinsically associated with the ULX. Based on the multiwavelength data,
we argue that the ULX is unlikely to be a background AGN. A more likely option 
is an accreting BH with a black hole mass of $\geq 10^4$\solmass, 
which might be a stripped remnant of a merging dwarf galaxy. 
\end{abstract}


\section{Introduction}
Ultraluminous X-ray sources (ULXs) are typically defined as bright and 
non-nuclear point sources with $L_{\rm X-ray} > 10^{39}$ erg s$^{-1}$, 
substantially larger than the Eddington 
luminosity of stellar mass black holes (but see \citealt{king_2001}). 
They have been proposed as candidates of intermediate-mass 
($10^2$\solmass $< M_{\rm BH} < 10^5$\solmass) black holes (IMBHs; 
\citealt{colbert_1999}). While IMBHs are a cosmologically important link 
between stellar mass black holes and supermassive black 
holes (\citealt{volonteri_2010}), they are very rare objects
(e.g., \citealt{greene_2004}). 
Thus, ULX could have great implication for studying the IMBH population.
However, ULXs are preferentially found in late-type or star-forming 
galaxies (e.g., \citealt{swartz_2004}). This might suggest that they originate 
from high-mass X-ray binaries (HMXBs; $20$\solmass $< M_{\rm BH} < 
100$\solmass; e.g., \citealt{gao_2003}), which might belong to the high-mass 
end of stellar BHs rather than IMBHs. Such massive stellar BHs are known
to form in low-metallicity environments (\citealt{mapelli_2013b}). In this
scenario, the power of ULXs could be explained either by accretion at 
super-Eddington rates (e.g., \citealt{begelman_2002})
or beamed emission (e.g, \citealt{king_2001}).
Thus, typical ULXs of moderate power ($L_{\rm X} < 2-5\times 10^{40}$
\lum) are occasionally regarded to be powered by HMXBs. However, ULXs exceeding 
this luminosity ($\geq 5-10\times 10^{40}$ \lum) are thought to imply 
supermassive or intermediate mass black holes at their center
(\citealt{swartz_2011}; but see \citealt{bachetti_2014}).

One of the most striking candidates for IMBH is a hyperluminous X-ray
source (HLX-1) in ESO 243$-$49 (S0 galaxy; \citealt{farrell_2009}). 
Its peak X-ray luminosity is  $\sim 10^{42}$ erg s$^{-1}$, which makes HLX-1 
the most luminous known ULX. It appears to have an optical counterpart. 
From the optical spectroscopy, \citet{wiersema_2010} found \ha\ emission,
whose central wavelength is consistent with the redshift of the host galaxy. 
Since the X-ray luminosity is substantially larger than the Eddington 
luminosity of HMXBs, the black hole mass is thought to be as
large as $\sim 2 \times 10^4$\solmass\ (\citealt{godet_2012}). 
While the origin of HLX-1 is unclear, its photometric properties 
indicate that it might be a remnant of the nucleus of a merging companion 
galaxy (\citealt{mapelli_2013a}).

Here, we present another ULX candidate IMBH.
This ULX (CXO J133815.6+043255) is found in the outskirts of NGC 5252. 
NGC 5252 is an early-type (S0) Seyfert 2 galaxy at a redshift of 0.0229. 
It is one of the best examples to present extended biconical emission-line 
region up to $\sim 30$ kpc from the nucleus (\citealt{tadhunter_1989}).
The ULX is located $\sim 10$ kpc away from the center. 
In this paper, we investigate the origin of the ULX in NGC 5252 using 
multi-wavelength data, including optical spectroscopic and UV/optical imaging 
data. 
The paper is organized as follows. In \S{2}, we describe the observing method
and procedures of data analysis. In \S{3}, we present the observed properties 
of the ULX based on the multi-wavelength data. In \S{4}, we discuss the 
nature of the ULX. 
Throughout the paper we adopt $H_0 = 71$ \kms\ Mpc$^{-1}$, $\Omega_m=0.27$, and 
$\Omega_\Lambda=0.73$, yielding a luminosity distance to NGC 5252 of 98.4 Mpc.  

\begin{figure*}[t]
\psfig{file=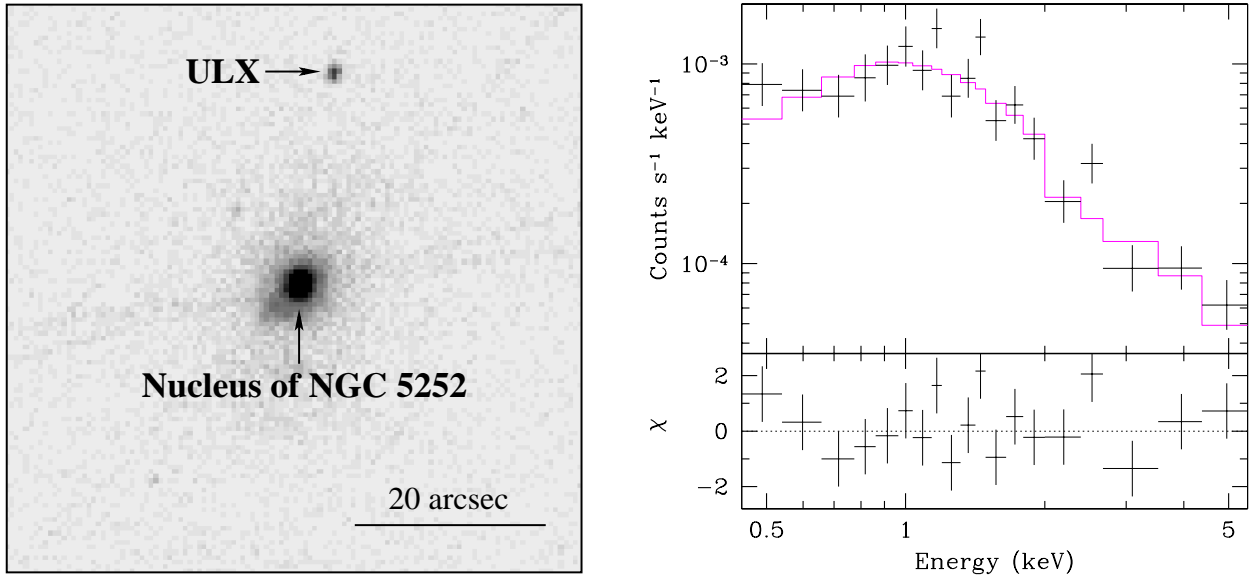,width=18.5cm}
\figcaption{
Left panel: Chandra ACIS-S image of the NGC 5252 in the 0.3--8 keV band, 
centered on the X-ray bright NGC 5252 active nucleus. The ULX is located to 
the north as indicated. 
North is up and east is left.
Right panel: X-ray spectrum of the ULX and the best-fit absorbed power-law 
model (magenta line). 
}
\end{figure*}

\section{Data}
\subsection{Chandra X-ray Observations}
NGC 5252 was first observed with the Chandra X-Ray Observatory 
(\citealt{weisskopf_2002}) on August 11, 2003 for 60.1 ks (observation 
identification number 
[ObsID] 4054, PI: M. Dadina), using the spectroscopic array of the Advanced 
CCD Imaging Spectrometer (ACIS) detector (\citealt{garmire_2003}) in the 1/4 
sub-array mode. The nucleus of NGC 5252 was placed near the aimpoint on the 
backside-illuminated S3 chip and the data telemetry mode was set to 
``Very Faint''. Three more ACIS-S observations with the same configuration were 
obtained on March 4, March 7 and May 9, 2013, with exposure time of 40.4 ks, 
67.8 ks and 62.5 ks, respectively (ObsIDs 15618, 15022, 15621; PI: J. Wang).  
The data were analyzed following the standard procedures using CIAO (Version 
4.6) with CALDB (Version 4.5.9) provided by the Chandra X-ray Center (CXC). We 
reprocessed the event files using CIAO  script {\tt chandra\_repro} to apply 
the latest version of relevant calibration. The light curve from regions free 
of bright sources on the S3 chip was used to identify time intervals of high 
background, and no further filtering was required for this data set. The total 
exposure time is 230.8 ks.

Although the Chandra observations were intended to image the diffuse X-ray 
emission in NGC 5252, a bright off-nuclear X-ray point source caught our 
attention besides the known X-ray bright Seyfert nucleus. Using CIAO tool 
{\tt wavdetect}, we have identified the source at a position of 
R.A.=13:38:15.64 and Dec.=+04:32:55.4. (Fig. 1).  
The uncertainty (1$\sigma$) on the source position is 0\farcs3.

\subsection{Spectroscopic Observations}
We obtained an optical long-slit spectrum using the Inamori-Magellan Areal 
Camera \& Spectrograph (IMACS) on the Baade 6.5 meter telescope at Las Campanas 
Observatory on May 29, 2013 UT. The data were taken with a 1\farcs5 slit 
in grism mode (300 lines/mm). It covers $3900-8000$\AA\ with a spectral 
resolution ranging from 650 to 1200. The sky was mostly clear but occasionally 
covered by thin cirrus clouds. The seeing was between 0\farcs6 to 1\farcs0. 
The observation was done in relatively low airmass ($<1.35$). The total 
integration time was 2 hrs, divided into 4 exposures to avoid the saturation
of the sky emission lines.
\psfig{file=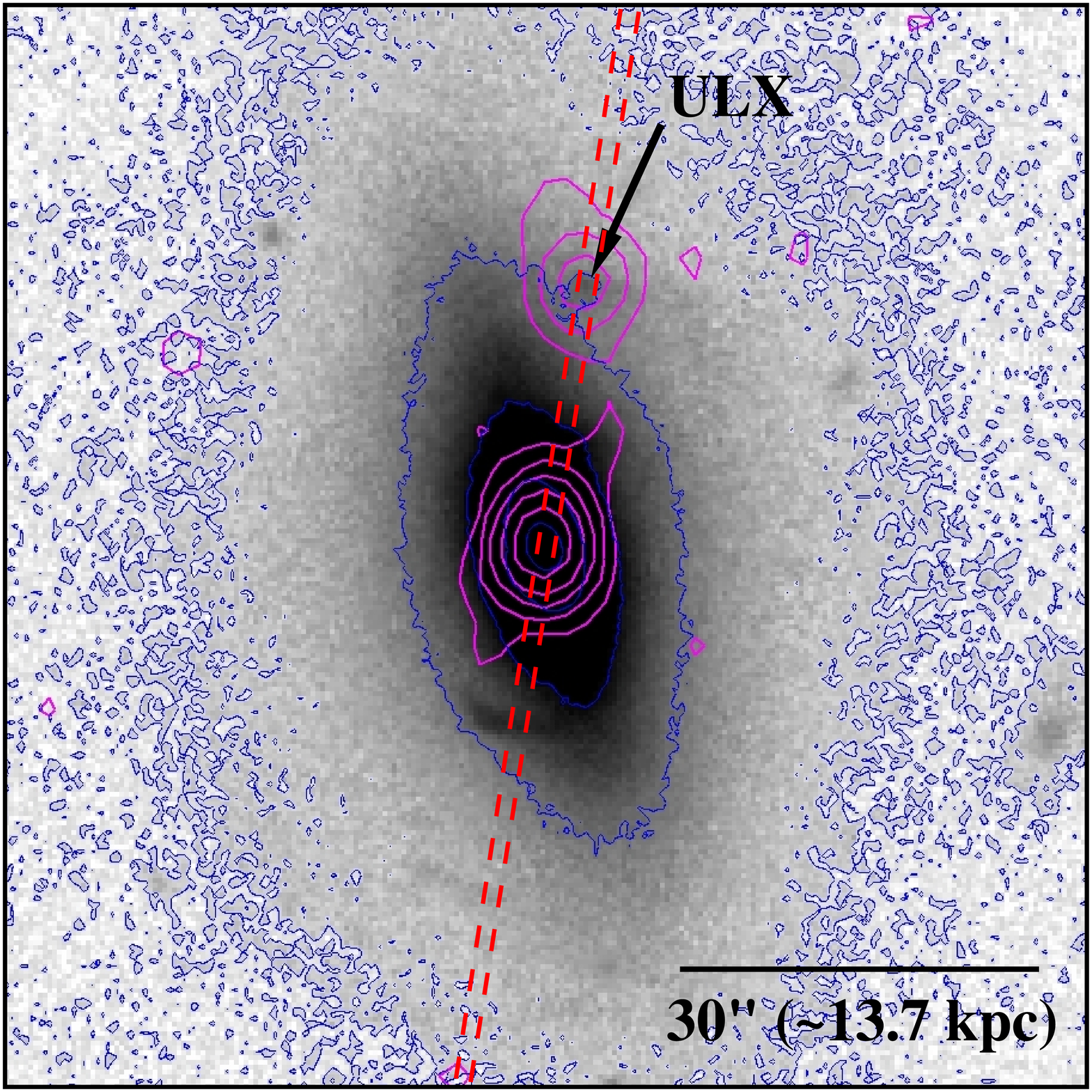,width=8.5cm}
\figcaption{
$g$-band image of NGC 5252 from Sloan Digital Sky Survey. 
Blue contours represent the surface brightness.
Magenta contours represent radio
continuum at 20 cm from the FIRST survey. 
The ULX appears to have optical and radio counterparts.
The long-slit position is denoted by the red dashed lines. 
North is up and east is left.
}
\vskip 0.2in
\noindent
The slit was aligned to cover the center of the galaxy and the ULX 
simultaneously so that we can gauge the contribution from the central AGN 
along the radius (Fig. 2). 
The slit ($\sim$30\amin) is long enough to cover the entire 
galaxy and the sky. The spatial pixel size is 0\farcs2 on the sky. 
Following the target observation, we took the arc spectrum 
and the flat frames for the calibration. The flux calibration was done with 
two standard stars (CD 32 and LTT 4816). 

\psfig{file=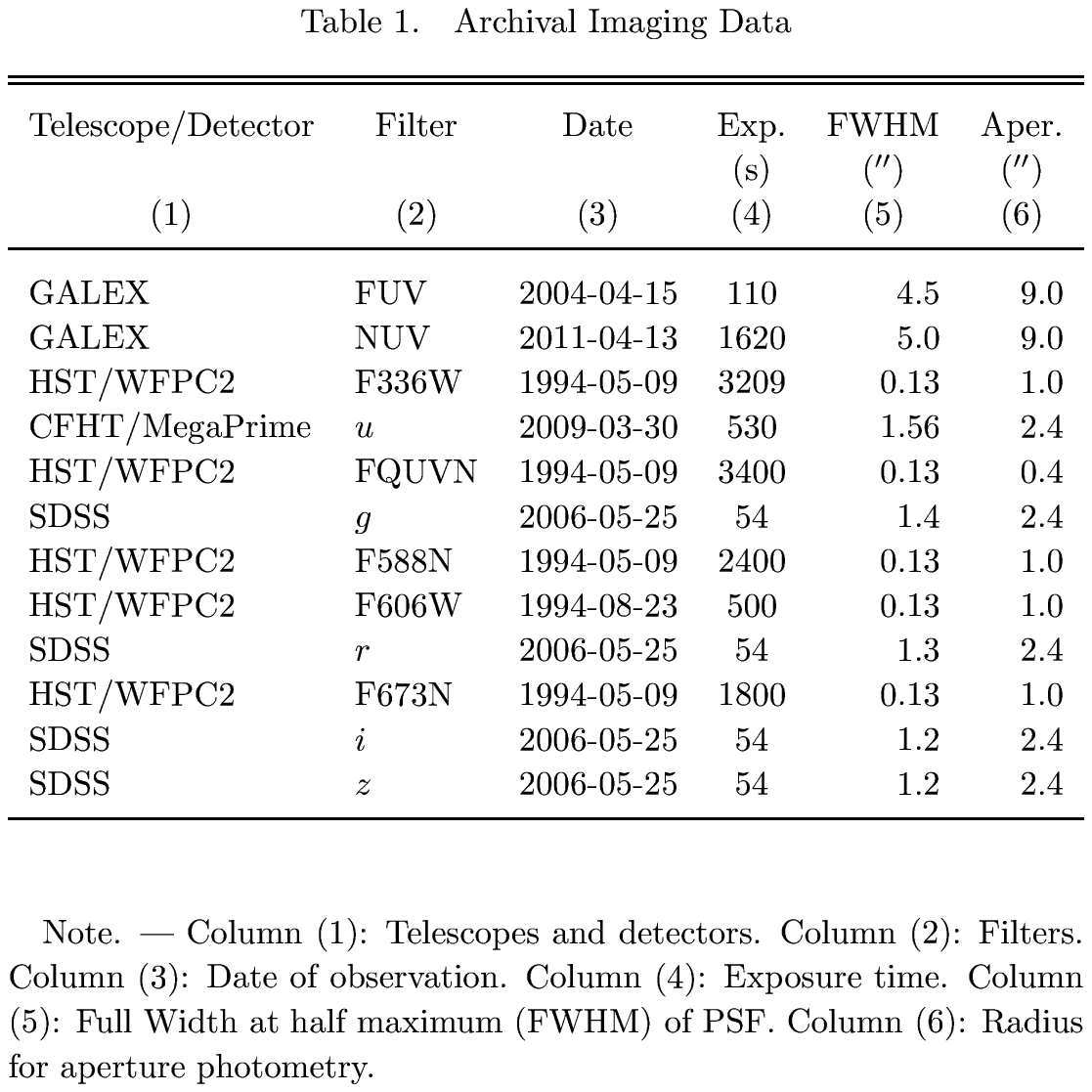,width=8.50cm}
\vskip 0.2in

We reduced the data using IRAF. Flat-field correction was done using 
two sets of images (dome flats and sky flats). We used He+Ne+Ar arc
lamp spectra for wavelength calibration. Spatial rectification was done
using negative features made by blocked gaps in every $\sim100$\asec\
along the long-slit in the slitmask (See Fig. 3). 
We combined the two-dimensional spectra using  cosmic ray
rejection, and applied the flux calibration. In order to see a spatial variance 
of the spectra, one-dimensional spectra with a width of 2\asec\ were extracted. 
Finally the spectra are corrected for Galactic extinction using the
extinction curve of \citet{fitzpatrick_1999}.

At the position of the ULX, we extracted a spectrum with 2\asec\ width and 
subtracted the local sky background in the vicinity of the source, so that we 
can remove the contribution from the galaxy. Throughout the paper, this 
spectrum is used for estimating the optical spectral properties of the ULX.

\subsection{Photometric Observations}
We assemble archival UV/optical imaging data taken with various telescopes 
[{\it Galaxy Evolution Explorer} (\galex), {\it Hubble Space Telescope} 
(\hst), Sloan Digital Sky Survey (SDSS), and Canada France Hawaii Telescope 
(CFHT)]. Table 1 summarizes the archival data used in this paper. 
The data include 12 images obtained with FUV, NUV, F336W, $u$, FQUVN,
$g$, F588N, F606W, $r$, F673N, $i$, and $z$ filters.
The basic data reduction steps of bias subtraction, flat-fielding,
and alignment are carried out by the pipeline of each dataset. 
For the data from \hst, we use the {\tt lacos\_im} task 
(\citealt{vandokkum_2001}) to remove cosmic rays. 
The flux calibration is also given by the pipeline, except the imaging data
from the CFHT. We indirectly derive the zeropoint of the CFHT/$u$ image by 
using the count rates of five bright stars with known magnitude from SDSS.

The ULX is visually detected in most images (Fig. 4). But, in FUV and $z$, 
it is unclear if the counterpart is present at the position of ULX. 
The position of the optical counterpart in the ground-based images is
well matched with that derived in X-ray data within its positional
\psfig{file=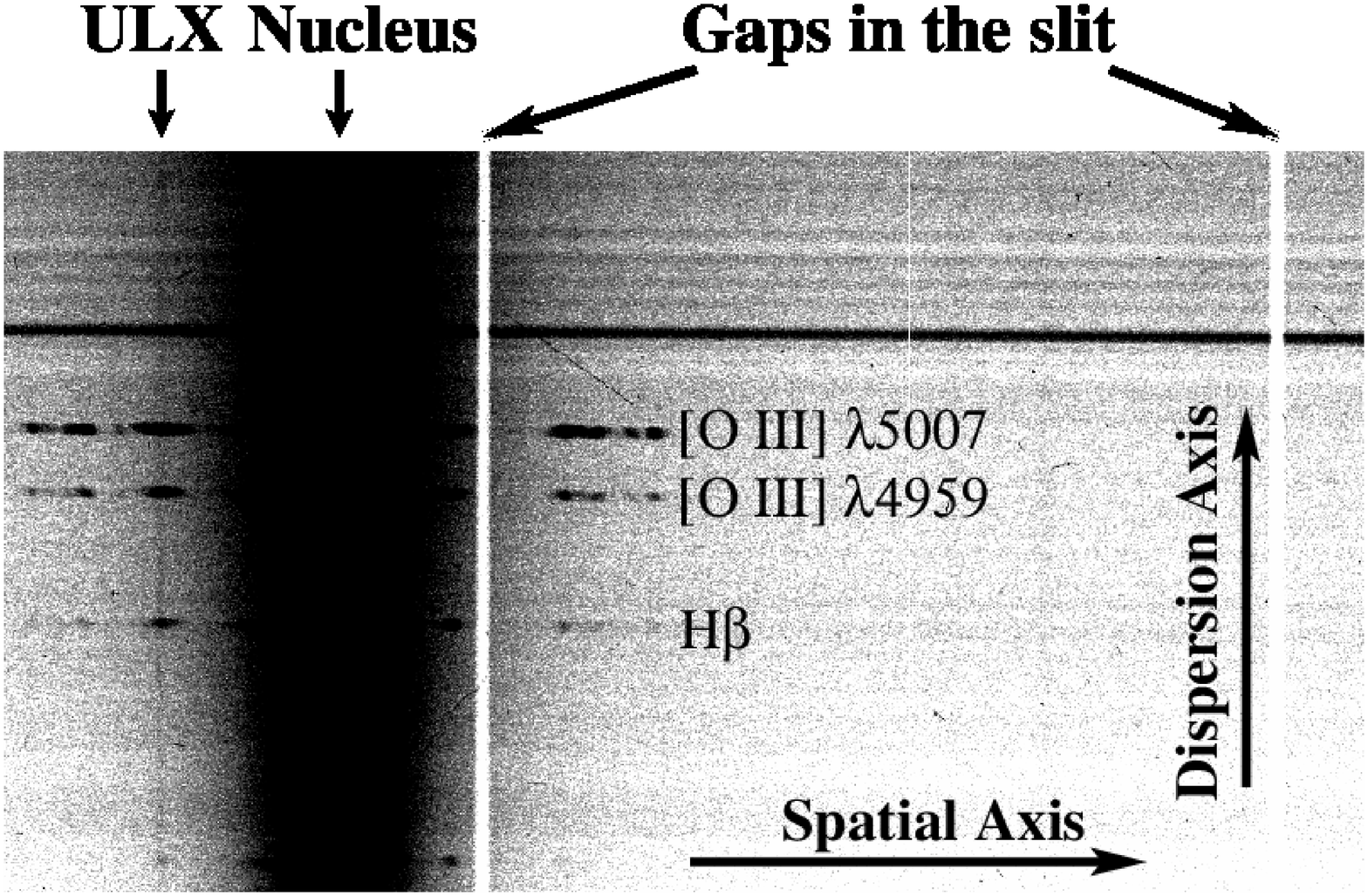,width=8.5cm}
\figcaption{
Part of two-dimensional spectrum of the nucleus of NGC 5252 and the ULX. 
The spatial extent is 170\asec. The positions of 
the ULX and the galaxy center are labeled. Continuum of the ULX 
is clearly seen. There are negative features along the dispersion axis, which
are generated by the gaps in the slit-mask. We use these features to 
calculate the distortion along the spatial axis. 
}
\vskip 0.2in
\noindent
uncertainty (0\farcs3). But the absolute astrometry of the \hst\ images,
especially those obtained before the Cycle 14, which have a typical astrometric
error of 
$1-3$\asec due to the uncertainty of the guide star positions 
(\citealt{koekemoer_2007}), is not as accurate as other data. 
Thus we correct the astrometry of the \hst\ images using the sources with 
known position from the SDSS images. However, due to narrow field-of-view 
and shallowness of the \hst\ images, we were only able to use three faint 
sources for the registration. Given the typical astrometric error of the SDSS 
images 
($\ge$ 0\farcs1; \citealt{pier_2003}), the uncertainties of the astrometry in 
the \hst\ images might be slightly larger than 0\farcs1. We again find that 
the optical counterparts in the \hst\ images are coincident with the X-ray 
position within 0\farcs5. Note that the aperture radii (0\farcs4 to 9\asec) 
for the aperture photometry are significantly larger than the astrometric 
uncertainties ($\sim$ 0\farcs1) of the optical images so that the photometric 
error due to the image misalignment should be negligible.

\begin{figure*}[t]
\psfig{file=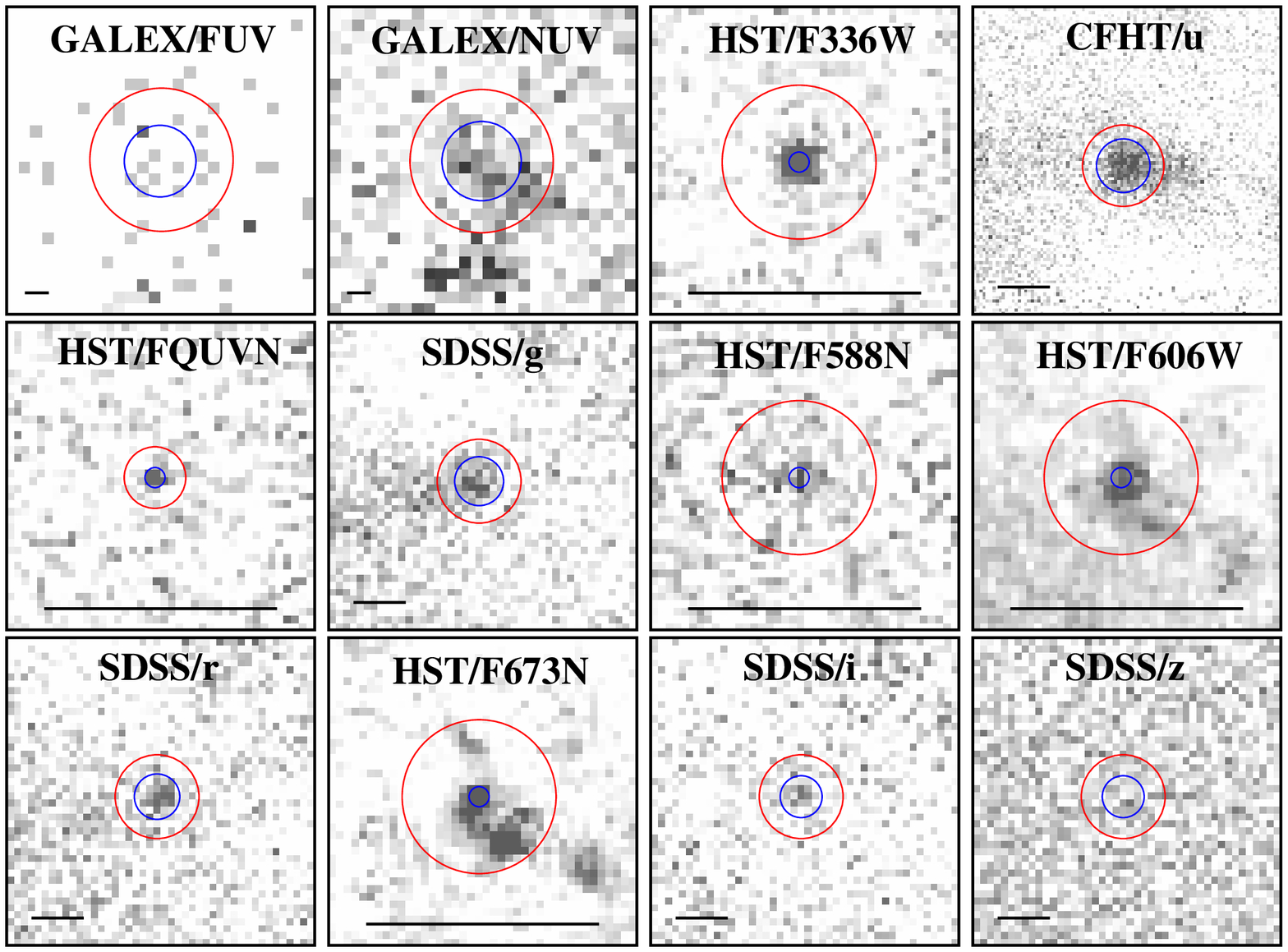,width=18.5cm}
\figcaption{
Postage stamp images of the ULX from archival data ({\it GALEX}, \hst, CFHT,
SDSS), listed in order of increasing wavelength. 
Inner blue circles represent twice the size (FWHM) of PSF. Outer red
circles denote radius for aperture photometry. 
The length of bar at the bottom of each image corresponds to 3\asec.
North is up and east is left for each image. All the sources are well matched 
with the position in the X-ray data within its astrometric uncertainty ($\sim$
0\farcs3). All images are presented in asinh stretch.
}
\end{figure*}

\section{Properties of the ULX}

\subsection{X-ray Properties}

We extracted X-ray counts in the energy range of 0.3--8 keV from a circular 
region of 2\farcs5 radius centered at the source. Background was taken 
locally from an annular region also centered at the source. The total number 
of background-subtracted source counts is $391\pm 21$. Spectra and instrument 
responses were generated and combined using {\tt specextract}. Spectra were 
grouped to have a minimum of 20 counts per energy bin to allow for $\chi^2$ 
fitting. The spectral modeling was performed with XSPEC Version 12.7 
(\citealt{arnaud_1996}). 
We fit the X-ray spectrum with an absorbed power-law model. The fit 
prefers negligible intrinsic absorption, so we fixed the absorption column at 
the Galactic neutral hydrogen column density in the sightline toward NGC 5252,
$N_{\rm H} = 1.97\times 10^{20}$ cm$^{-2}$ (given by \citealt{dickey_1990}). 
The best fit power-law photon index $\Gamma=1.65 \pm 0.11$, with a reduced 
$\chi^2$/dof=19.8/17. Adding in an extra thermal emission or power law 
component does not further improve the fit statistically. The absorption 
corrected 0.5-8 keV flux is $1.33\pm 0.11 \times 10^{-14}$ erg 
cm$^{-2}$ s$^{−1}$.  Adopting a distance of 98.4 Mpc to NGC 5252, this 
corresponds to $L_{\rm X}=1.5\times 10^{40}$ \lum, placing 
the source 
into the ULX domain. We further examined the light curves for 
intra-observation and inter-observation variabilities, and found no 
significant variations. 
\psfig{file=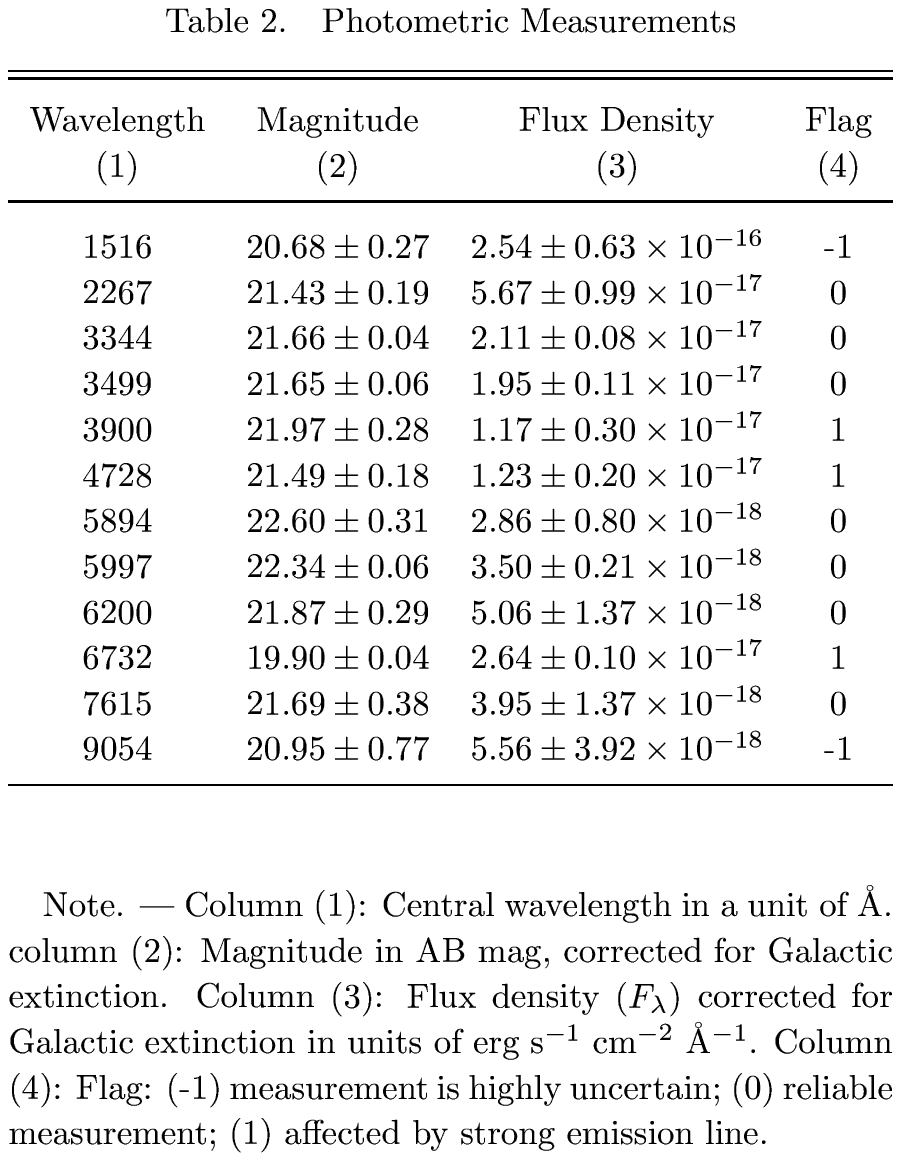,width=8.5cm}
\vskip 0.2in
\noindent
Comparing to a constant light curve, the probability 
of variable signal is negligible in all cases. However, we found a long-term
variability over 10 years, evaluated using the count rates in 2003 and 2013.  
A $\chi^2$ test against a constant light curve indicates the probability for 
no variation is 0.0002, which implies the long-term variation is statistically 
significant.

\begin{figure*}[t]
\psfig{file=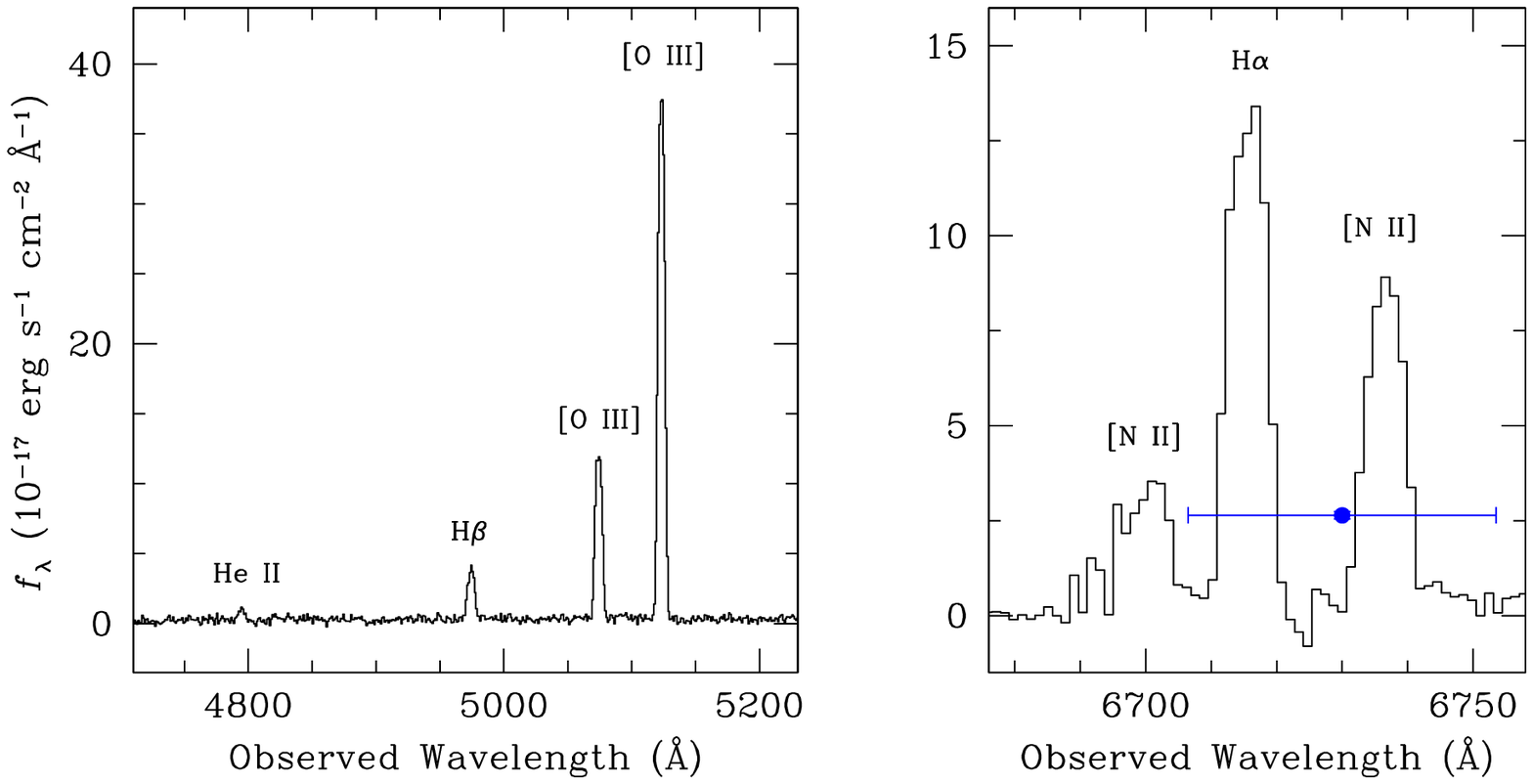,width=18.5cm}
\figcaption{
Optical spectrum at the position of the ULX extracted with an aperture width 
of 2\asec. In the right panel, we overplot the photometric measurement 
(blue circle) from the \hst\ image obtained with the F673N filter. 
}
\end{figure*}

\subsection{Spectral Properties}

Strong narrow emission lines are found in the position of the ULX (Fig. 5).
The \oiii\ and \ha\ luminosities corrected for 
the extinction using the observed Balmer decrement (\ha/\hb$\approx3.9$) are 
as large as $10^{39.7}$ and $10^{39.2}$ \lum, respectively.
We note that these fluxes might be overestimated by 0.2 dex due to the 
extended structures in the vicinity of the ULX (\S{3.3}).
Thus if the lines are ionized by the source in the ULX, its energy is 
comparable to that of nearby low-luminosity AGNs (\citealt{ho_2008}).   
However, it could also be ionized by the central AGN, which is 
$\sim22$\asec\ away from the ULX.

Figure 6 shows distributions of \oiii\ flux and systematic velocity along the 
slit. As already known from previous studies, NGC 5252 harbors an 
extended narrow-line region (ENLR) up to a radius of 50\asec\ 
(e.g., \citealt{tadhunter_1989}). The ULX is located on the major 
axis of the ENLR. \citet{acosta_1996} also reported an excess of line flux 
in \oiii\ and \ha\ close to the ULX, while the slit in their study is slightly 
off from the ULX (NW2 in \citealt{acosta_1996}). 
Using the flux ratios among optical emission lines  (\hb, \oiii\ \lamb 5007, 
\ha, \nii\ \lamb 6583), one can determine the source of ionizing radiation
(\citealt{baldwin_1981}). 
Figure 7 indicates that the ENLR is mainly ionized by 
the AGN rather than hot stars from star-forming region, which is in
good agreement with previous studies (e.g., \citealt{acosta_1996}). 
Since the strong X-ray emission, optical emission lines, and radio continuum 
can be associated with shocks (e.g., \citealt{sulentic_2001}; 
\citealt{konstantopoulos_2013}),
it is worth testing this possibility. But the flux ratios among oxygen lines 
[log (\oiii\ \lamb\lamb4363/5007) $\approx-1.9$ and log 
(\oii\ \lamb3727/\oiii\ \lamb5007) $\approx-1.6$] lead us to rule out the
a significant contribution from the shocks (\citealt{dopita_1995}; see also 
\citealt{morse_1998}).

Due to the low instrumental spectral resolution 
($\sigma_{\rm ins}\sim 100-200$ \kms), the emission lines are barely spectrally
resolved at the position of the ULX. The velocity distribution shows a 
rotating pattern within $\sim5-6$\asec\ of radius from the nucleus, which is 
in a good agreement with previous studies (\citealt{tsvestanov_1996}; 
\citealt{morse_1998}). But above that radius, the dynamics of the ionizing gas 
appears to be complicated. Interestingly, we note that the systematic velocity 
offset of the ULX relative to the nucleus of the galaxy is fairly small 
($\sim 13$ \kms).
This indicates that the ULX might be gravitationally bound to NGC 5252.

\psfig{file=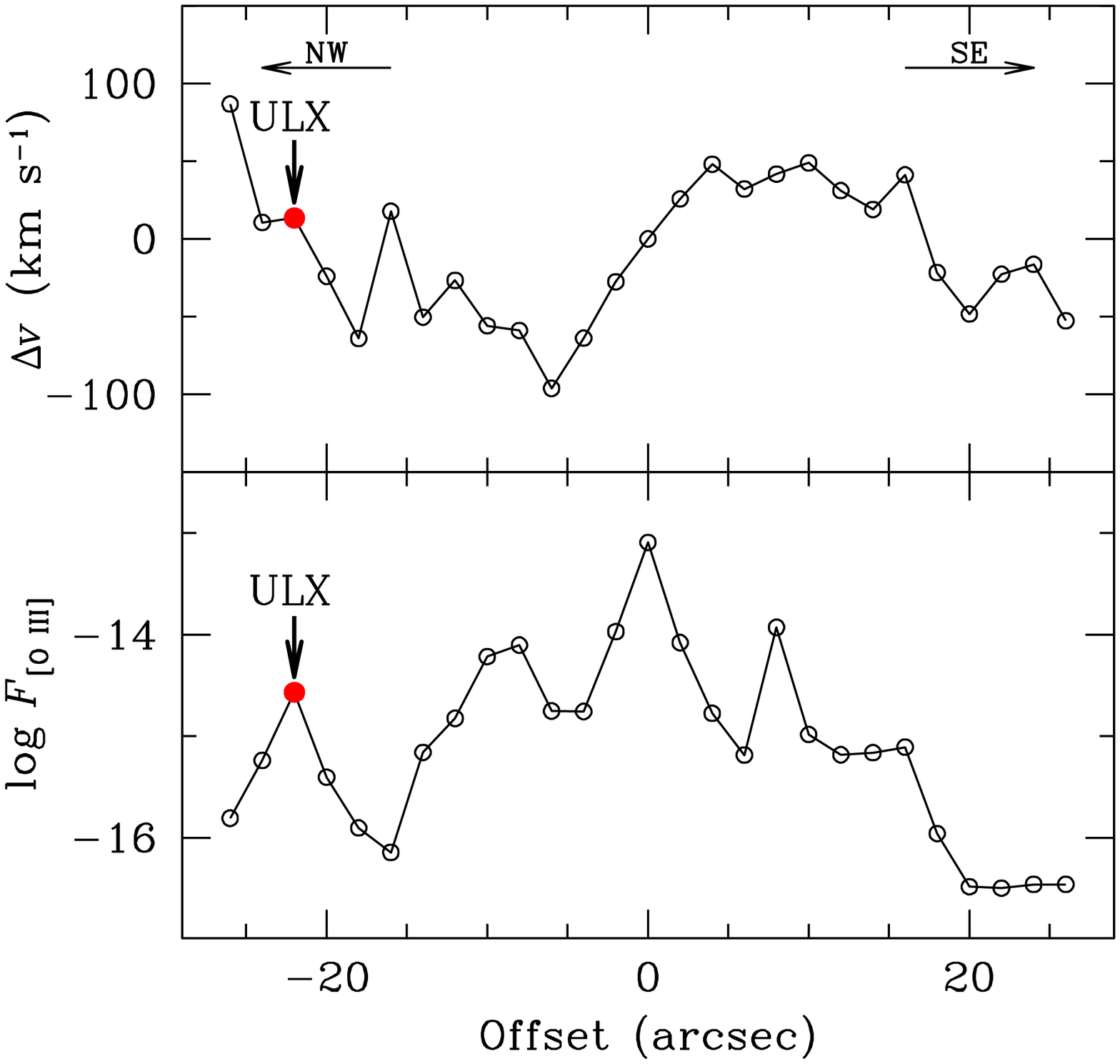,width=8.5cm}
\figcaption{
Distributions of \oiii\ \lamb 5007 flux in units of erg s$^{-1}$ cm$^{-2}$
(bottom), and systematic velocity relative to the center (top), estimated from 
the peak of \oiii\ \lamb 5007, as a function of radius. 
The ULX is located $\sim 22$\asec\ northwest of the center (red filled circle).
}
\vskip 0.2in

Besides the strong emission lines, we find weak sign of underlying continuum
(see Fig. 3). The S/N of the continuum flux is $\approx2.0$ between 
$\sim5000$\AA\ and $\sim8000$\AA, while the S/N dramatically decreases 
($\leq 1$) below 4500\AA. Thus the absolute flux calibration below 4500\AA\ 
may be very uncertain. Whereas the source of the continuum is unclear, the 
spectral energy distribution (SED) based on the
broadband photometry suggests that it is not entirely explained by the stellar
continuum unless the stellar population is very young 
($\sim10^{6-7}$ yrs). 
No broad component is detected in \hb\ or \ha.
We carefully examine the 
\psfig{file=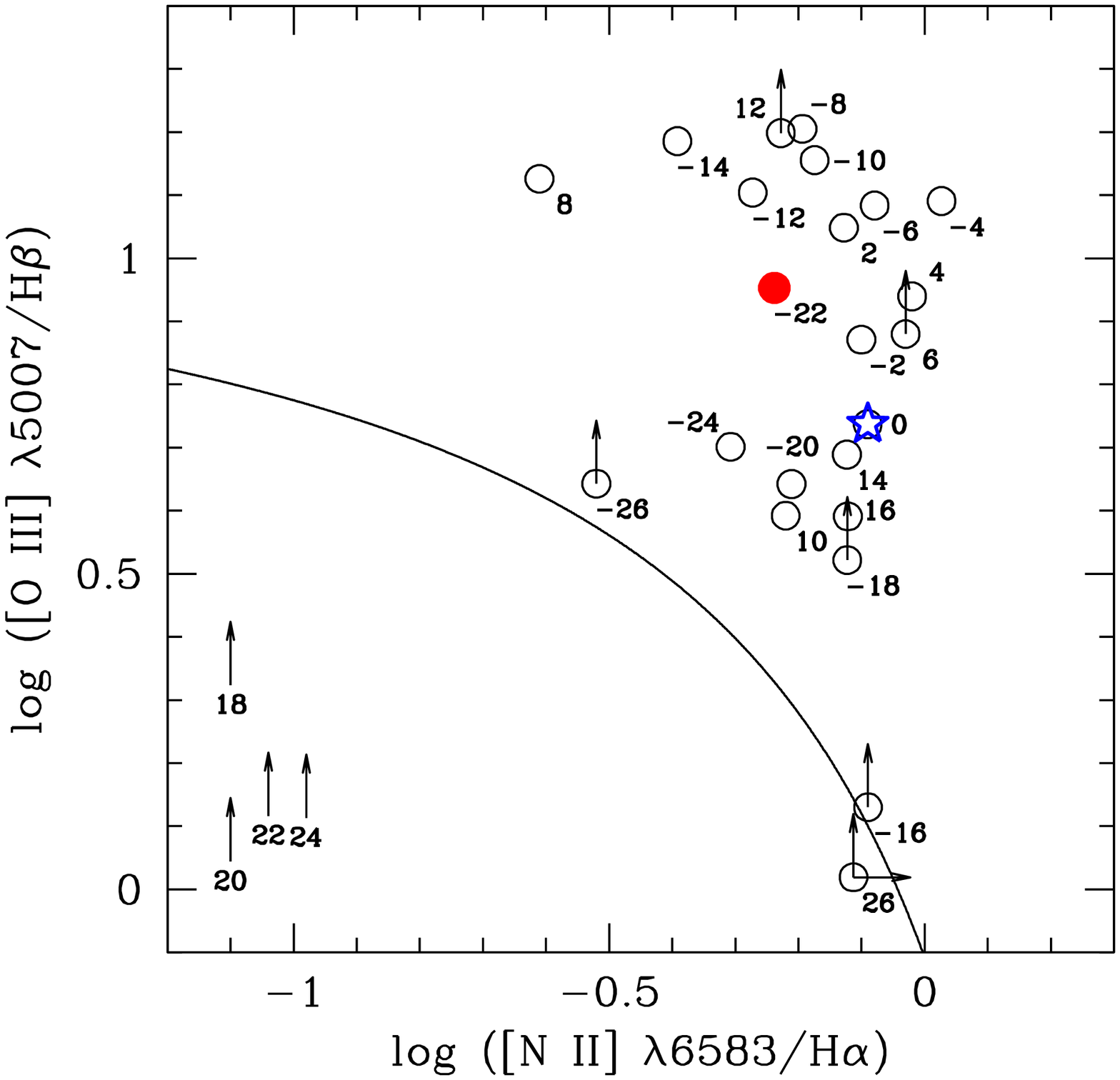,width=8.5cm}
\figcaption{
Distributions of optical line ratios for various positions. The distance from 
the center for each position in units of arcsec, which is same as the offset
in Figure 6, is denoted below the point. 
The center
of NGC 5252 and the ULX are denoted by a blue star and a red filled circle, 
respectively. Data with no measurement of \ha\ and \nii\ is plotted in the
bottom-left corner. The solid line shows classification line between
AGNs and star-forming galaxies, derived by \citet{kewley_2001}.
}
\vskip 0.2in
\noindent
spectrum of the ULX to test the hypothesis that X-ray 
continuum comes from a background source.  But we fail to find any signature 
(e.g., emission line, featureless continuum) of its existence. We will discuss 
this further in \S{4.2}. 

\subsection{Photometric Properties}
The high-resolution images from \hst\ reveal that the optical counterpart is 
compact and marginally resolved (e.g., F336W), while there is sign of extended 
features associated with the ULX in the low-resolution images 
(e.g., NUV, $u$, and $g$) and the \hst\ images
with narrow-band filters affected by strong emission lines (e.g., F673N).
Using the F336W image, which has the highest S/N among the images, we estimate
the size of the optical counterpart. Using GALFIT 
(\citealt{peng_2002}; \citealt{peng_2010}), we fit the source by 
employing a synthetic point-spread function (PSF) generated by TinyTim 
(\citealt{krist_1995}). We model the source with a \ser\ (\citealt{sersic_1968})
profile or a Gaussian profile. The best fit yields that the effective radius
($r_e$) is $\sim0$\farcs1 (46 pc). However, since, the derived $r_e$ is 
comparable to the size of the PSF, thus shall be regarded as 
an upper limit. Even with this caveat, it is intriguing that this upper limit 
of the size is consistent with that of ultracompact dwarfs but significantly 
larger than that of globular clusters (e.g., \citealt{forbes_2013})

Since the archival images are obtained with different detectors and in 
different observing conditions, it is important to measure the fluxes of
the ULX in a consistent way. Moreover, it is crucial to remove the 
light of the host galaxy correctly because the gradient in the background can 
introduce an additional error on the photometric measurements.
We model the host galaxy component with a \ser\ profile using GALFIT. We find 
that the host galaxy is well fit with 2 or 
3 \ser\ components, but we do not ascribe physical significance to each 
individual component. 
After the host galaxy contribution is removed, we perform aperture 
photometry as a function of radius. The optimal aperture radius (red
circles in Fig. 4) is determined where the cumulative flux within the radius 
reaches a 
constant. If the source appears to be extended (e.g., $\it u$ and F673N),
we use the aperture radius determined from other images obtained with the 
same or similar detector. Table 2 lists the flux 
measurements from the aperture photometry. While the flux in the FUV is 
very uncertain ($\sim$ 0.3 mag uncertainty), 
we note that it is also contained in the \galex\ 
archival catalog (\citealt{morrissey_2007}); the catalog lists an AB magnitude
of 20.58 mag, which is slightly 
brighter than, but within the uncertainty, our measurement of 20.92 mag. 
Thus, our measurement in FUV is reliable.

\begin{figure*}[t]
\psfig{file=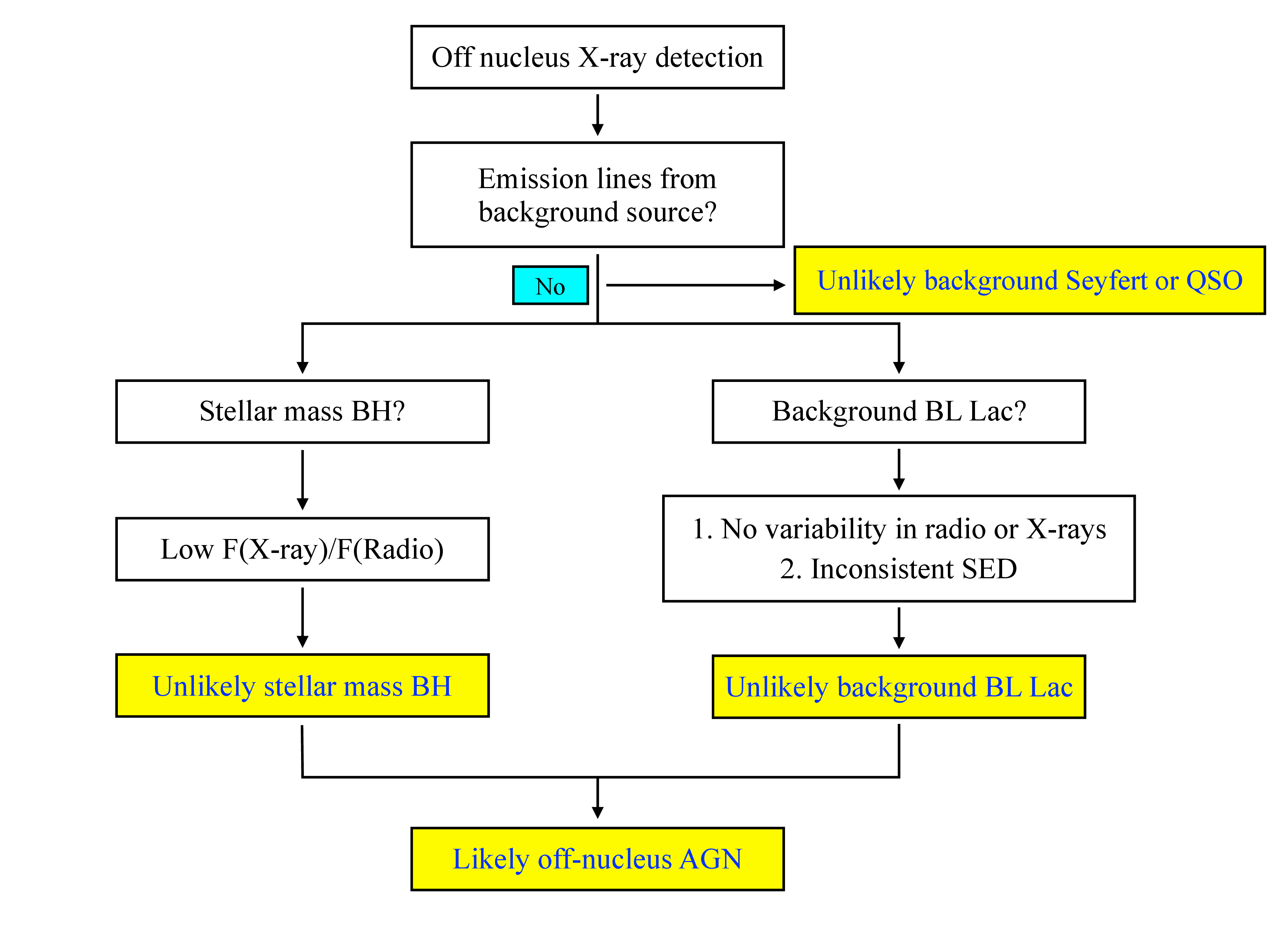,width=18.5cm}
\figcaption{
Flow chart for diagnosis of the nature of the ULX based on the observed 
photometric and spectroscopic properties. 
}
\end{figure*}
 
The F673N and F606W images show that there is faint, extended 
substructures to the southwest and northeast of the ULX, probably arising 
from \ha$+$\nii emission.
This emission likely lies partly or entirely within the aperture size of our
extracted spectrum, although its nature and its association with the ULX are
unclear.  From aperture photometry of the F673N and F606W images with and
without the substructures masked, we estimate that its contribution to the ULX
spectrum is at most 40\%. Therefore, it is possible that the luminosities of
emission lines (e.g., \oiii\ and \ha) are overestimated by 0.2 dex due to the 
substructures. Throughout this paper, we take this factor
into account.

The optical counterpart is also detected by SDSS Data Release 7 
(\citealt{abazajian_2009}). Their measurements are broadly in good agreement 
with ours in $g$, $r$, and $i$. But, there is substantial discrepancy in $u$ 
and $z$, which are much shallower than other bands. For the $u$ band, we use 
a deeper image obtained with CFHT rather than that from SDSS. The flux from 
the $z$ band is quite uncertain with an error of $\sim0.77$ mag. 

We note that the flux density from the spectroscopy broadly agrees with 
the broadband photometry, except the blue part below 4000\AA\ (Fig. 10). 
While it 
is unclear why the discrepancy is remarkable in the blue part, it indicates 
there might be variability in the UV. This finding may be regarded as further 
evidence that the continuum source is nonstellar. 
Because we subtracted the local host galaxy light in the vicinity of the ULX 
when we extracted the spectrum (\S{2.2}), the contribution from the host 
galaxy should be negligible.

\subsection{Radio Data}
Radio images of NGC 5252, obtained multiple times for $\sim3$ yrs 
(from 1991 to 1993), show a source in coincidence the position of the ULX 
(\citealt{wilson_1994}; \citealt{kukula_1995}; \citealt{becker_1995}).
The offset between the radio peak and the X-ray position of the ULX is smaller 
than 0\farcs3.
The radio flux densities of the ULX are on average $1.50\pm0.12$, 
$1.9\pm0.2$, and $3.14\pm0.06$ mJy at $3.6$, $6$, and $20$ cm, respectively. 
While the beam sizes of the radio images range from 0\farcs1 to 5\asec,
the radio source at the position of the ULX is unresolved for all the radio 
data. The spectral index 
of the radio continuum between 3.6 and 20 cm of the counterpart of the ULX
is $\sim$0.5 ($\alpha$ in $f_\nu \propto \nu^{-\alpha}$), which is consistent
with that of nearby Seyferts (\citealt{ho_ulvestad_2001}).
We find that the 
variability over 3 years is less than 10\% at 3.6 and 20 cm (see also 
\citealt{wilson_1994}). 

\section{Nature of the ULX}
We presented the spectroscopic and photometric properties of the optical 
counterpart of the ULX, obtained from the various telescopes and instruments.
Based on this dataset, we characterize and discuss its nature. The flow chart 
in Figure 8 briefly summarizes our approach to 
understand the origin of the ULX.

\subsection{The ULX is likely to be associated with NGC 5252}
Strong emission lines are detected at the position of the ULX and its velocity
offset with respect to the center of NGC 5252 is remarkably small ($\sim 13$ 
\kms). This suggest that the ULX is associated with NGC 5252. However, it is 
still inconclusive mainly because the emission lines
underneath the ULX could be contamination from the ENLR. To address this issue, 
we investigate the profile of \oiii\ in the vicinity of the ULX. Intriguingly, 
we find a velocity gradient within a radius of 2\asec. As shown in Figure 9, 
the velocity inferred from the peak value of the \oiii\ line is maximized at 
radii of $\sim -2$\asec\ and 1\farcs4 (916 pc and 640 pc), where the maximum 
velocities are $-80$ and 120 \kms, respectively. 
This strongly suggests that the line-emitting gas is affected by and therefore 
is physically associated with or in close proximity to the ULX.  If so, then 
the ULX is not a background source but is physically located within NGC 5252.

While the \oiii\ line is barely resolved, it appears that the line dispersion 
slightly varies ($\Delta \sigma \leq 50$ km s$^{-1}$) within a radius of 
2\asec, which indicates that the kinematics of the gas are affected by 
the ULX.  This further suggests that the ULX is located within NGC 5252. 
The gas itself may be physically excited by the ULX, which appears to be 
powered by an active BH.

\subsection{Unlikely to be a Background AGN}
Although we do detect emission lines at the position of the ULX, it is still
difficult to prove that the line emission comes from the ULX or is, instead, 
simply confused by the ENLR.  Thus, it is worthwhile to consider whether the 
ULX might be an interloper.
It is common that ULXs turned out to be foreground stars or 
background AGNs (\citealt{foschini_2002}; \citealt{masetti_2003}; 
\citealt{gutierrez_2005}; \citealt{wong_2008}; 
\citealt{gutierrez_2013}). Moreover, the incidence of background AGNs 
appears to be higher for ULXs found in early-type galaxies (e.g., 
\citealt{Gutierrez_2006}; \citealt{heida_2013}). 

As discussed in \S{3.2}, we do not see any signs of emission lines, narrow or 
broad, from a background AGN in the spectrum.  How strong should any 
hypothetical line emission be?
We use the empirical correlation between X-ray and \oiii\ luminosity to 
estimate the strength of the expected optical line emission, if the source 
were a line-emitting background AGN. We adopt a conversion 
factor of ${\rm log}\ (L_{\rm [O\ III]}/10^{42} \ {\rm erg\ s^{-1}}) = 
0.91\ {\rm log}\ (L_{\rm 2 keV}/10^{42} \ {\rm erg \ s^{-1}}) - 1.21$ 
based on type 1 AGNs (\citealt{stern_2012b}; \citealt{stern_2012a}) 
\psfig{file=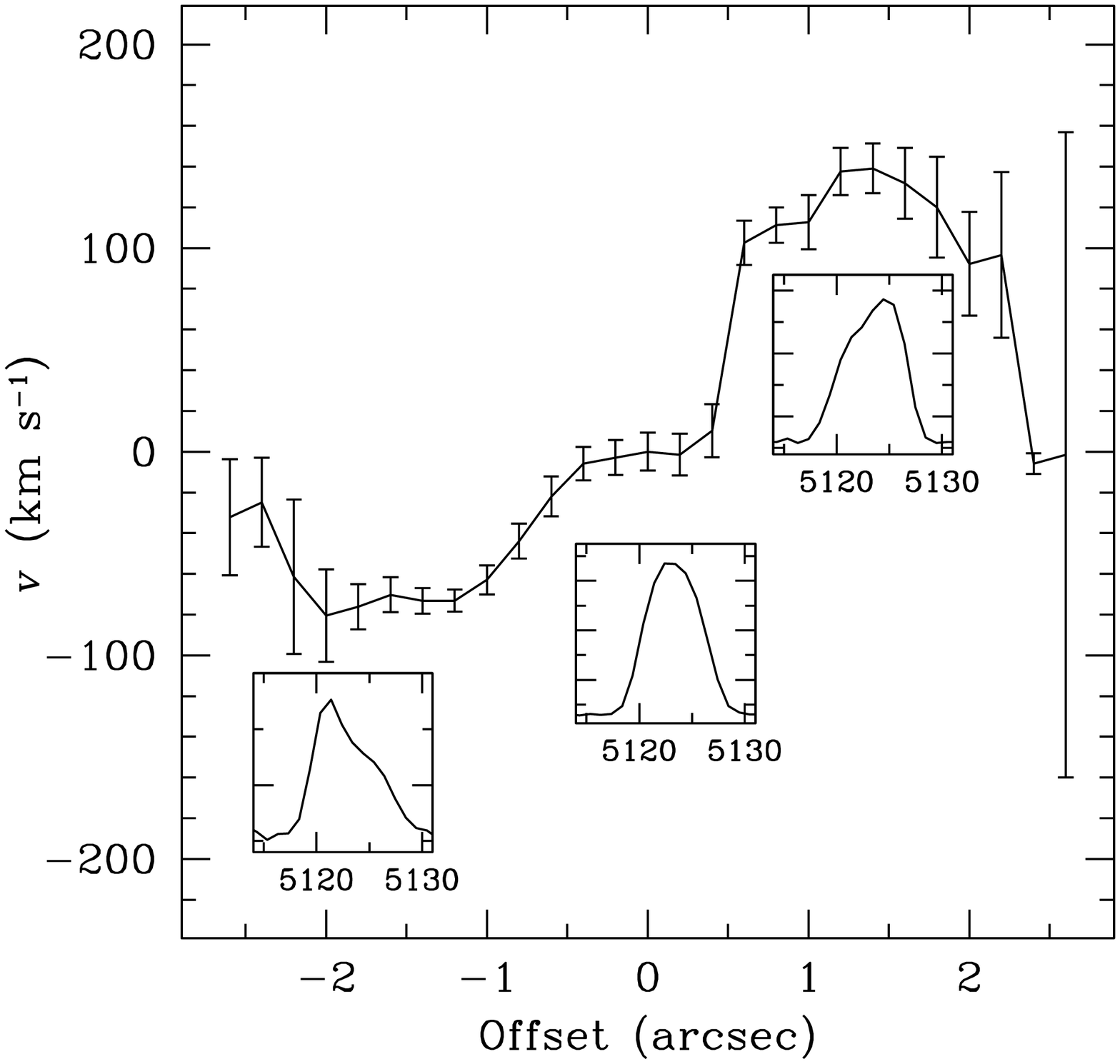,width=8.5cm}
\figcaption{
Velocity distribution derived from the peak of \oiii\ emission as a function
of radius from the center of the ULX. From left to right inner panels,  we plot 
profiles of \oiii\ at -2\asec\ (NW), 0\asec, and 1\farcs4 (SE) away from the 
center, respectively.
}
\vskip 0.2in 
\noindent
to calculate 
the expected \oiii\ luminosity, given a rest-frame 2 keV luminosity estimated from the observed $\Gamma= 1.65$.
\oiii\ can be detectable up to $z\sim 0.8$ due to the wavelength limit of the 
spectrum. We expect an \oiii\ luminosity ranging from $10^{40}$ to 
$1.5\times10^{42}$ erg s$^{-1}$ at $0.1<z<0.8$. 
Given the S/N of the continuum of the optical spectrum, we can detect \oiii, 
if it is unresolved at the instrumental resolution of our spectrum, up to 
z $\sim 0.8$ at a significance of 6$\sigma$. This estimate assumes 
negligible internal extinction, which is not unreasonable given the low 
observed X-ray column.

Having ruled out the possibility that the ULX is a low-redshift AGN, we now 
evaluate whether it can be an AGN at z $> 0.8$.  
With $L_{\rm 2keV} > 10^{43.5}$ 
\lum, such a hypothetical source would be luminous enough to qualify as a Type 
2 quasar (e.g., \citealt{ptak_2006}), but the low observed column density is 
incompatible with the heavy obscuration typical of this class 
(\citealt{jia_2013}).  At the same time, the featureless optical spectrum, 
which would correspond to rest-frame UV, is also inconsistent with that of a 
type 1 quasar.  To estimate the expected strength of broad UV emission lines, 
we adopt the empirical scaling between the strength of the \oiii\ and Mg II 
and C IV, as well as the median line width of FWHM = 4500 km s$^{-1}$, as 
derived from the large SDSS quasar catalog of \citet{shen_2011}.  Given the 
S/N of our spectrum, we can rule out the existence of broad UV emission lines 
at the level of 10$\sigma$.

From the above analysis, we reject the possibility that the ULX is a 
line-emitting background AGN at any reasonable redshift.  What about a 
weak-line AGN, namely a BL Lac object?  We argue against this possibility 
using three lines of evidence.
The compactness of the radio source (\citealt{kukula_1995}) is consistent
with the radio morphology of BL Lacs. 
First, the SED of the ULX does not look like that of typical BL Lacs 
(\citealt{fossati_1998}; \citealt{donato_2001}). Figure 10 (left) shows that 
there is substantial discrepancy, especially in the UV, between the SED of the 
ULX and those of BL Lacs.  We varied the redshift but could not find any 
satisfactory solution that can simultaneously match the observed SED of the 
ULX in the UV and X-ray bands.  One obvious weakness in this conclusion, of 
course, is that the SED measurements were not simultaneous, whereas BL Lac 
objects are known to be highly variable.

Second, the optical properties of the ULX do not agree with the expected 
properties of a BL Lac host. BL Lac objects are commonly hosted by massive 
(e.g., \citealt{falomo_1996}; \citealt{urry_1999}; \citealt{kotilainen_2005}),
luminous ellipticals with $M_R = -22.9 \pm 0.5$ mag (e.g., 
\citealt{sbarufatii_2005}). Thus if the ULX
is a BL Lac object, one can expect that the host galaxy would be detected in 
the \hst\ images. By using the distribution of the apparent $R$-band 
magnitude of BL Lac host galaxies (\citealt{sbarufatii_2005}), we find
that the host galaxy would have been detected if the ULX is a low-redshift BL
 Lac with $z<0.4$. However, a high-z BL Lac can not be ruled out as an 
interloper because the expected r-band host brightness for BL Lacs with 
$z \geq 0.4$ would be lower than our observational limits.

\begin{figure*}[t]
\psfig{file=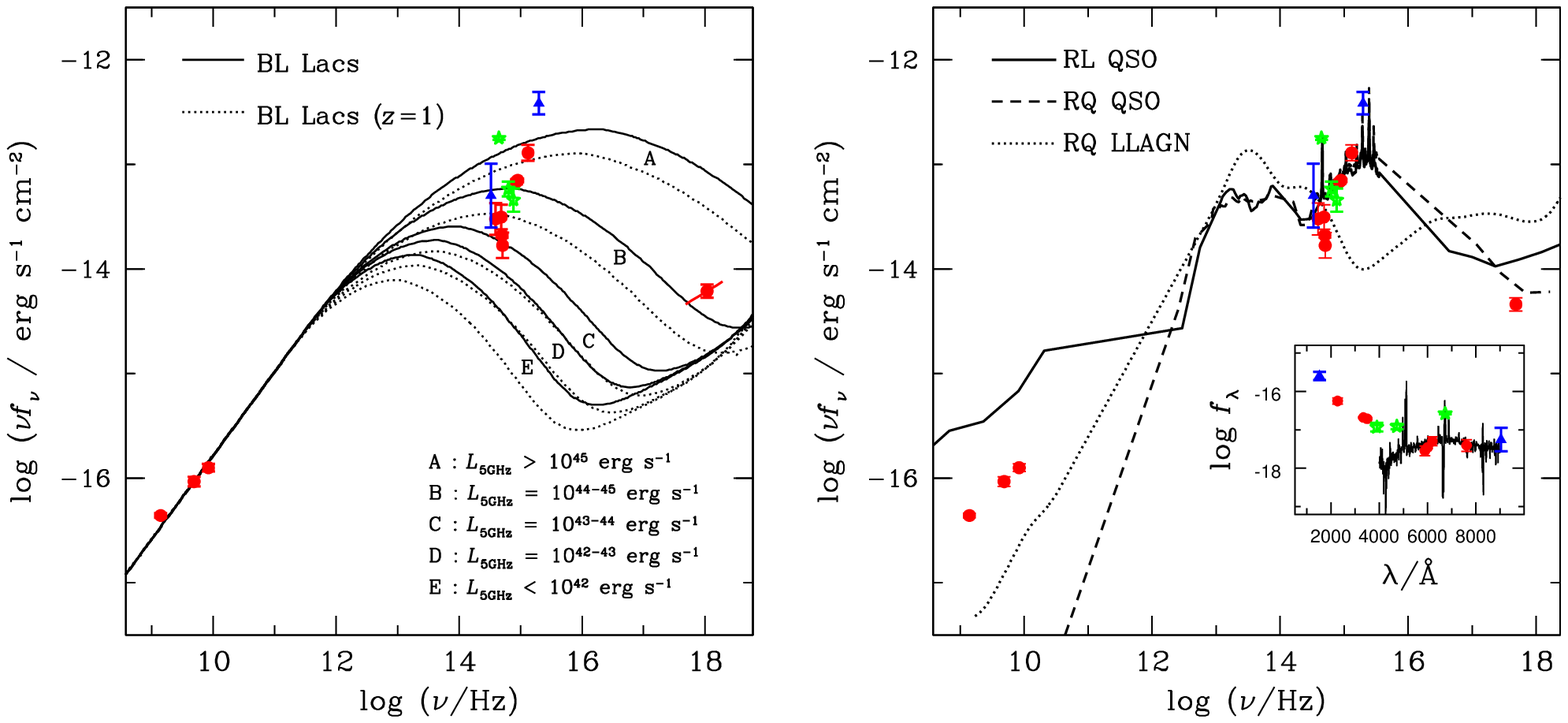,width=18.5cm}
\figcaption{
Spectral energy distribution (SED) of the ULX, assuming that the ULX is 
located in NGC 5252. 
Ambiguous measurements are denoted by blue triangles. Photometric measurements 
affected by strong emission lines are denoted by green stars. Red circles 
represent robust measurements.
({\it left}) Solid lines represent template SEDs of BL Lacs as a function
of their radio luminosity (log $L_{\rm 5GHz}$) from \citet{donato_2001} 
(see also \citealt{fossati_1998}). 
Dotted lines represent the same SEDs of BL Lacs but redshifted to $z=1$. 
They are normalized to that of the ULX at 5 GHz. 
({\it right}) Solid line and dashed line denote SEDs  of 
radio-loud (RL) and radio-quiet (RQ) quasars, respectively, from 
\citet{shang_2011}.
Dotted line shows SED of radio-quiet low-luminosity AGNs 
($L_{\rm bol}/L_{\rm Edd} < 10^{-3.0}$) from \citet{ho_2008}.
SEDs of quasars and low-luminosity AGNs are normalized to that of the ULX
at 6000\AA.
In the inner panel, we overplot the optical spectrum of the ULX for 
comparison. The spectrum is smoothed using a Gaussian kernel for the 
purpose of display. Note that, for the purpose of display, the units of the 
inner panel are different from those of the larger panel.
}
\end{figure*}



Finally, it is noteworthy that the ULX source does not vary in the radio by 
more than 10\%.  BL Lacs, by contrast, are highly variable ($> 10\%$) in the 
radio (\citealt{aller_1992}). However, it is possible that we do not have 
enough 
data to correctly judge the variability because the radio fluxes were obtained 
in only three different epochs within 3 years.  
In order to test this possibility, we perform a Monte Carlo simulation by 
randomly choosing the data points at three different epochs from monitoring 
radio data of BL Lacs (\citealt{richards_2011}). We find that $\sim40\%$ of
BL Lacs show variability less than $10\%$. This indicates that additional
multi-epoch data are needed to use radio variability to test whether the ULX 
is a background BL Lac. Interestingly, the X-ray data measured at 
three different epochs for 5 days also do not exhibit short timescale 
variability at a level greater than 17\%.   

It is very intriguing that the SED of the ULX is roughly consistent with that 
of radio-loud quasars (\citealt{shang_2011}; see Figure 10, right panel). 
However, the lack of emission lines, as discussed above, is strongly at odds 
with this interpretation.
\\
\subsection{Not a stellar-mass BH}
As we discuss above, the probability that the ULX is a background
AGN appears to be relatively low. Under the assumption that the X-rays, 
optical emission lines and radio continuum are associated with the ULX, it 
is natural to suppose that it is an off-nucleus stellar mass BH. For stellar 
mass black hole binaries, the X-ray flux is correlated closely with the radio 
flux (e.g., \citealt{corbel_2003}; \citealt{gallo_2003}). Using the empirical  
correlation from \citet{gallo_2003}, we find that the expected
radio flux density ($\sim0.003 - 0.022$ mJy) is 2 orders of magnitude smaller 
than the observed flux value ($\sim1.5 - 3.1$ mJy) if the ULX is a `low/hard' 
state black hole binary. 
For an X-ray binary to be as luminous in the radio as observed relative to the 
X-rays, it would have to be in the ``very high'' state 
(\citealt{fender_2004}).  
However, this is incompatible with the lack of X-ray and radio 
variability, which is typically observed in very-high state X-ray binaries
(\citealt{miyamoto_1991}; \citealt{kubota_2004}).
In addition, the X-ray spectrum is expected to be very steep in the very high 
state ($\Gamma \geq 2.5$; \citealt{miyamoto_1991}; \citealt{hori_2014}), 
which is inconsistent with 
the value of $\Gamma=1.65$ observed for the ULX.  We thus argue that our 
source is unlikely to be a stellar-mass BH.


\vskip 0.2in
\subsection{Off-nucleus non-stellar BH}
If the ULX is an off-nucleus non-stellar BH, for example from the accretion of
a tidally stripped galaxy nucleus, 
one can expect that the line emission should correlate with the ionizing 
continuum from the BH. To test this scenario, we use the well-known 
correlation between \oiii\ luminosity and
X-ray luminosity in AGNs (e.g., \citealt{bassani_1999}; 
\citealt{heckman_2005}).
Figure 11 shows that the observed $L_{\rm X}/L_{\rm [O III]}$ of the ULX is 
comparable to that of local Seyferts. The X-ray luminosity from 2 to 10 keV is 
corrected for Galactic and intrinsic absorption. The \oiii\ luminosity is 
corrected
for extinction using the observed Balmer decrement (\ha/\hb$\approx3.9$) and
the extinction curve of \citet{fitzpatrick_1999}. 
As discussed in \S{3.3}, the \oiii\ luminosity may have been overestimated 
by 0.2 dex. Thus, a lower limit of the \oiii\ luminosity is also displayed in 
Figure 11; however, this is not significant enough to change any of our 
conclusions.
That the ULX follows the $L_{\rm X}$ $-$ $L_{\rm [O III]}$ relation of 
Seyferts suggests that the optical line emission is powered by the local 
X-ray continuum rather than the central AGN in the 
host galaxy.
As expected, the nucleus of NGC 5252 also falls on the relation; the X-ray and 
\oiii\ luminosities are estimated 
within the nucleus region 
(\citealt{cruz_gonzalez_1994}; \citealt{dadina_2010}). Since the X-ray and
\oiii\ emissions are broadly extended, the luminosities might be slightly 
underestimated. But the amount of underestimation should be negligible because
the flux contribution from the extended region is less than 1\% 
(\citealt{dadina_2010}). 

A correlation between radio power and line (\oiii) luminosity in bright AGNs 
has been reported by various studies (e.g.,  \citealt{rawlings_1989}; 
\citealt{miller_1993}). This relation appears to hold also for Seyferts 
(\citealt{ho_peng_2001}). We find that the observed luminosity ratio between 
\oiii\ {($\approx 10^{39.5-39.7}$ \lum)} and radio continuum at 
5 GHz ($\approx 10^{21.3}$ W Hz$^{-1}$) is in broad agreement with Seyferts 
(Fig. 11). Indeed, the ULX is bright enough in the radio to be classified as a 
radio-loud AGN. These findings suggest that the radio, X-ray continuum, and
\oiii\ emission originate from the same source, namely an off-nucleus 
non-stellar, presumably fairly massive BH. 

\begin{figure*}[t]
\psfig{file=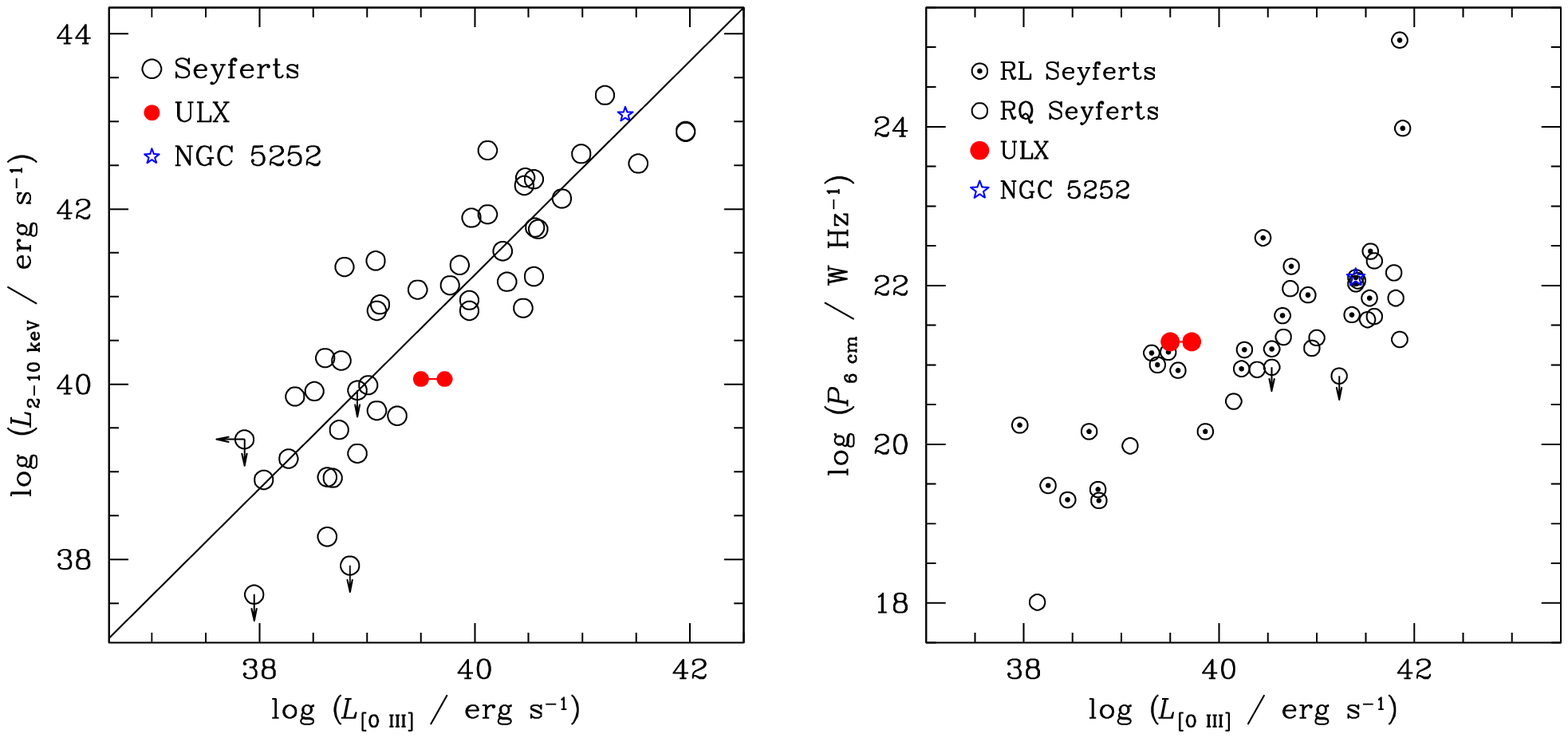,width=18.5cm}
\figcaption{
(Left) Correlation between \oiii\ and X-ray luminosity.
Open circles and red filled circle show local Seyferts from 
\citet{panessa_2006} and the ULX from this study, respectively. The solid 
line represents the relation obtained from the local Seyferts. 
The nucleus of NGC 5252 is denoted by a blue star. The range of \oiii\ 
luminosities plotted for the ULX accounts for the uncertainty introduced by 
the effect of masking its nearby extended substructures (Fig. 4). 
(Right) Correlation between \oiii\ and radio luminosity at
5 GHz. We plot radio-loud and radio-quiet Seyferts from 
\citet{ho_peng_2001}. Typical uncertainties of the \oiii\ and radio 
luminosities are 0.3 dex and 5\%, respectively. 
}
\end{figure*}
\vskip 0.2in

If the ULX is an off-nucleus BH, what is its mass? In order to address this 
question, we investigate the physical properties of the ULX. 
In the hypothesis that the bolometric luminosity does not exceed the 
Eddington luminosity, we can estimate a lower limit of 
the BH mass from the observed luminosity (but see \citealt{abramowicz_1988}). 
The bolometric 
conversion factors for the \oiii\ and X-ray luminosity strongly depend on
luminosity, in the sense that they 
($L_{\rm bol}/L_{\rm [O III]}$ and $L_{\rm bol}/L_{\rm 2-10 keV}$) tend to be
smaller for more low-luminosity AGN (e.g., \citealt{ho_2008}; 
\citealt{stern_2012b}). Adopting the conversion equation from 
\citet{stern_2012b}, $L_{\rm bol}/L_{\rm [O III]} \sim 162$, which leads to 
{$ M_{\rm BH} > 10^{3.5}$\solmass}.  We can 
carry out the same experiment using the X-ray luminosity. By adopting the 
bolometric conversion factor of $16-28$ for low- and moderate- luminosity AGNs 
(\citealt{ho_2008}), we estimate \mbh $\geq 10^{3.0-3.6}$\solmass, 
consistent with the estimate based on \oiii\ luminosity.

The BH mass can be also independently estimated using the fundamental plane 
of BH accretion (\citealt{merloni_2003}), which is described as 
\begin{equation}
{\rm log}\ L_{\rm R} = 0.6 {\rm log}\ L_{\rm X} + 0.78 {\rm log}\ 
M_{\rm BH},
\end{equation}
where $L_{\rm R}$ is the luminosity at 5 GHz in units of erg s$^{-1}$,
$L_{\rm X}$ is the luminosity at 2-10 keV in units of erg s$^{-1}$,
and $M_{\rm BH}$ is the BH mass in units of \solmass.
By assuming that the ULX follows this relation, we find that \mbh\ is 
as large as $\sim 10^8$\solmass\ (Fig. 12). 
\psfig{file=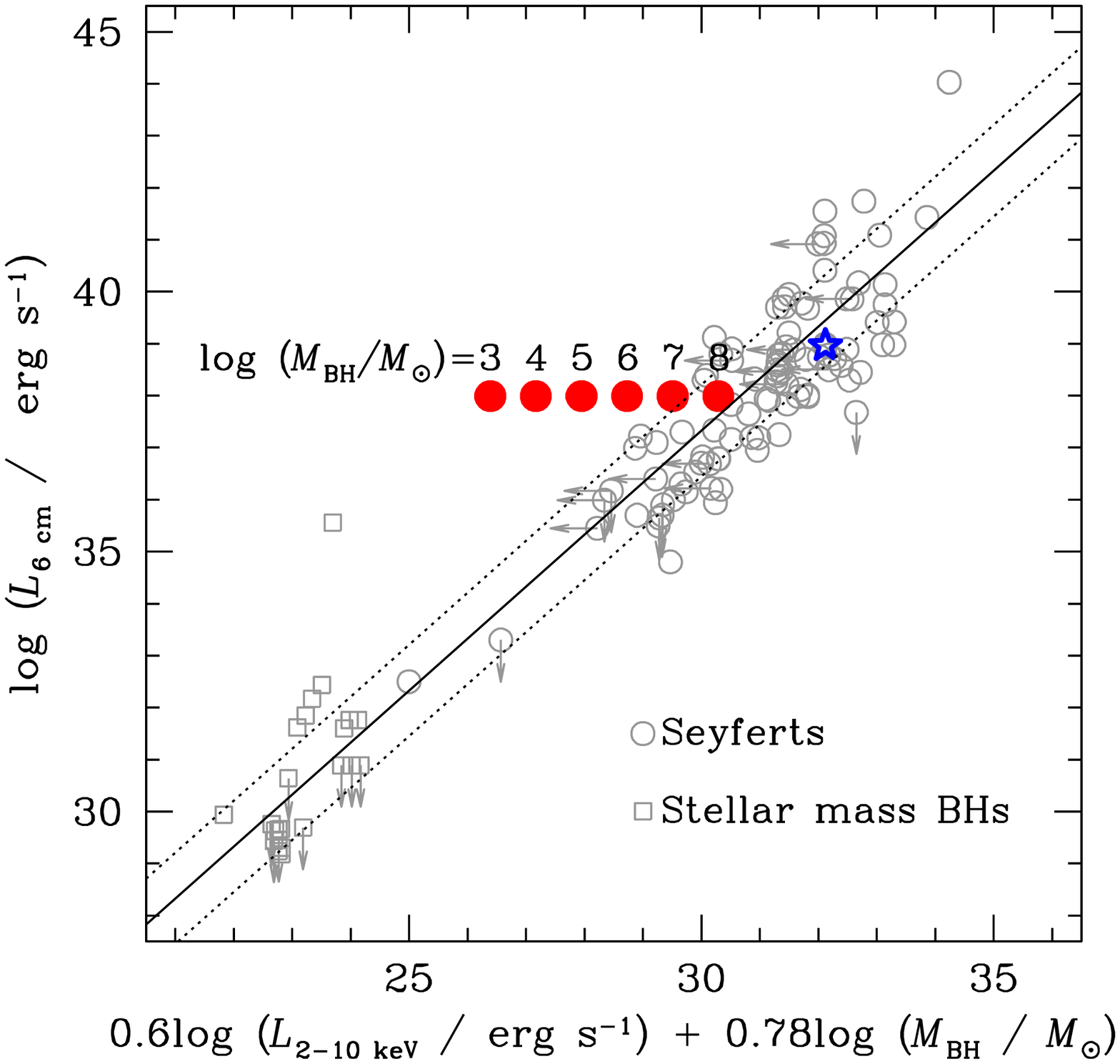,width=8.5cm}
\figcaption{
Fundamental correlation among radio luminosity, X-ray luminosity, and
BH mass. Open grey circles and squares show the relation for local Seyferts 
and stellar mass BHs, respectively (\citealt{merloni_2003}). The solid and 
dotted lines are the
best fit and its 1$\sigma$ scatter obtained by \citet{merloni_2003}, 
respectively. Blue star represents the center of NGC 5252 
($M_{\rm BH}=1.3\times10^8$\solmass, inferred from a relation between BH 
mass and stellar velocity dispersion). 
Red filled circles
show expected locations in the plane as a function of BH mass 
($10^{3-8}$\solmass) for the ULX. 
}
\vskip 0.2in
\noindent
While the fundamental plane has been recalibrated with various types of 
sample (e.g., \citealt{gultekin_2009}; \citealt{plotkin_2012}), our estimate 
does not change significantly if we adopt the revised versions.
A caveat is that this empirical relation is only applicable for BHs accreting 
in the low state (i.e. $L_{\rm bol}/ L_{\rm Edd} \leq 0.1$). 
If the BH mass is indeed as large as suggested by the fundamental plane 
relation, we are faced with a serious challenge.  First, the Eddington ratio 
would be very low ($\sim 2 \times 10^{-5}$).  Such a low Eddington ratio 
should produce a low-ionization parameter for the optical emission lines 
(\citealt{ho_2008,ho_2009}), similar to those seen in low-ionization 
nuclear emission-line 
regions, which have \oiii/\hb\ $< 3$ (\citealt{ho_1997}).  Instead, the ULX is 
observed to have Seyfert-like line ratios, with \oiii/\hb\ $\sim 9$.

If the ULX is really associated with a non-stellar mass BH, it raises 
the question about its nature. \citet{morse_1998} showed that the kinematics of 
the ionized gas in NGC 5252 can be modelled with at least three different 
components, which suggests an external origin of the gas. Although NGC 5252 is 
an early-type galaxy, it has abundant \hi\ gas 
($M_{\rm HI}\sim10^9$ \solmass), and its structure is highly complicated 
(\citealt{prieto_1996}).
Thus it seems that NGC 5252 may have undergone a gas-rich merger. 
In this case, we might speculate that the off-nucleus BH powering the ULX is the remnant of an already-merged companion galaxy.
If our estimation for the BH mass ($\sim 10^{4-8}$\solmass) is correct, the 
merging galaxy should have had a bulge with stellar mass of 
$\sim 10^{7.0-10.4}$\solmass, derived from the correlation between BH mass 
and bulge stellar mass (\citealt{kormendy_2013}). 

For comparison, we estimate the total stellar mass of NGC 5252 to be 
$M_*\approx1.4\times10^{11}$\solmass, using its $K$-band luminosity
($M_K=-24.55$ mag; \citealt{peng_2006}) and stellar velocity dispersion
($\sigma_*=209$ \kms; \citealt{cid_2004}) following the conversions given in 
\citet{kormendy_2013}. 
If the ULX has a BH mass as large as $\sim10^8$\solmass, its progenitor 
bulge of $M_*\sim10^{10.4}$\solmass\ implies that NGC 5252 experienced a fairly 
significant merger with a mass ratio of at least 1:6.  This seems highly 
implausible, in view of the fact that NGC 5252 shows no evidence for any 
perturbation (e.g., tidal features, lopsidedness, etc) in its current stellar 
distribution.   
On the other hand, it is 
not unreasonable that NGC 5252 could have accreted a small dwarf galaxy with 
$M_*\sim10^7$\solmass\ that would have accompanied a $\sim10^4$\solmass\ BH.  
Such a minor merger event should have little impact on the global structure 
of a galaxy $10^4$ times more massive.

Unlike most ULXs studied to date, this particular ULX is unusual in that it 
has a prominent counterpart in the optical continuum. The physical nature of 
the optical continuum is unclear.  We do not detect any obvious stellar 
features, but it is difficult to rule out that the continuum comes mainly from 
starlight, presumably the nucleus of a larger galaxy whose outer parts have 
already been stripped.  In the following, we {\it assume} that the optical 
continuum is stellar and use it to place an upper limit on the stellar mass 
associated with the ULX. To minimize possible
contribution from the featureless continuum of the BH accretion disk, we use
$i$-band photometry, which gives $M_i=-13.3$ mag. While mass-to-light ratio 
strongly depends on the initial mass function, metallicity, and age of stellar 
population, we approximately estimate an upper limit of stellar mass of 
$\sim 10^{8}$ \solmass\ using the stellar population synthesis models from 
\citet{vazdekis_2012} and \citet{ricciardelli_2012}. 

 
Interestingly, the properties of the ULX are comparable to those of a 
ultracompact dwarf galaxy (M60-UCD1) around NGC 4649 (\citealt{strader_2013}).
M60-UCD1 is very compact ($r_e \approx 24$ pc) and one of the most luminous 
UCDs known ($M_g \approx -13.7$). There appears to be an X-ray source at the 
center of M60-UCD1, although its luminosity ($\sim 10^{38}$ \lum) is 
significantly lower than that of the ULX studied here. While the origin of the 
X-ray emission is 
somewhat uncertain, it is very likely to originate from a massive BH at 
the center, rather than a stellar BH (\citealt{sivakoff_2007}; 
\citealt{strader_2013}; \citealt{seth_2014}).    
\citet{strader_2013} and \citet{seth_2014} argue that M60-UCD1 
is a tidally stripped remnant of a merging galaxy, which is consistent with 
our expectation for the ULX in NGC 5252. 

\subsection{Comparison with HLX-1 and NGC 3341}
Interestingly, the properties of the ULX in this study are somewhat 
similar to those of HLX-1 in ESO 243-39. Both of them seem to have 
BH significantly larger than stellar mass BHs, reside in early-type 
galaxies at similar distances ($z\sim0.023$), and have optical counterparts. 
In addition, their origin is thought to be merging galaxies. However, 
at the same time, it is intriguing to note that there are significant 
differences between the two sources. The ULX in NGC 5252 shows much higher flux 
ratio of optical emission to X-ray continuum.  The luminosity of \ha\ emission 
of the ULX in NGC 5252 is $\sim40-90$ times greater than that of HLX-1
(\citealt{wiersema_2010}), whereas the X-ray luminosity of the ULX in NGC 5252 
is only $1-65$\% of that of HLX-1 (\citealt{servillat_2011}). This implies 
that the difference in $L_{\rm X}/L_{\rm H\alpha}$ between the two sources is 
a factor of $90-9000$. Moreover, the radio flux density of the ULX in NGC 5252
is at least $\sim40$ times larger than that of HLX-1 (\citealt{webb_2012}). 
This might suggest that they have distinctive physical properties (e.g., 
BH mass and/or accretion rate). 

NGC 3341 is another interesting case of an off-nucleus source exhibiting
AGN signatures both in the optical and X-rays 
(\citealt{barth_2008}, \citealt{bianchi_2013}). Optical images of 
NGC 3341 clearly show evidence of ongoing merging with multiple companions
(\citealt{barth_2008}). The off-nucleus AGN appears to be associated with a 
dwarf galaxy, indicating that this AGN might be triggered during  
ongoing minor mergers. 

\section{Summary}
We report the serendipitous discovery of a ULX located 10 kpc from the 
type 2 Seyfert nucleus of NGC 5252, an S0 galaxy known to contain an extended 
narrow-line region.  The ULX has an intrinsic 2-10 keV X-ray luminosity of 
$L_X = 1.2 \times 10^{40} $ \lum, and the X-ray spectrum can be well-fit with 
a simple power law with a photon index of $\Gamma = 1.65\pm0.11$ with 
negligible intrinsic absorption column.  
No X-ray variability is detected within the observation window of 5 days.

The ULX has counterparts in the radio, optical and UV bands.  Follow-up 
optical spectroscopy reveals strong narrow emission lines characteristic of 
photoionization by an accretion-powered source.  The redshift of the emission 
lines coincides precisely with the systemic velocity of NGC 5252.  We argue 
that the \oiii\ emission, which is well correlated with both the X-ray and 
radio luminosity in the same manner as other AGNs, is intrinsically associated 
with the ULX, and that the emission in all bands arise from an accreting BH 
with a mass of at least $10^4$\solmass, located within NGC 5252.  Using a 
variety of arguments based on the optical spectrum and the broad band SED, 
we dismiss the possibility that the ULX is associated with a background AGN.  
We also consider improbable that the ULX is an X-ray binary system.

With a BH mass of $\geq 10^4$\solmass, the progenitor host of the ULX appears 
to have been a small dwarf galaxy $10^4$ times less massive than NGC 5252, 
which has since been accreted and assimilated into the larger galaxy.

\acknowledgements
We are grateful to an anonymous referee for helpful comments.
We thank John Mulchaey and Elina Nieppola for useful suggestions and 
discussion. LCH acknowledges support by the Chinese Academy of Science through 
grant No. XDB09030102 (Emergence of Cosmological Structures) from the 
Strategic Priority Research Program and by the National Natural Science 
Foundation of China through grant No. 11473002.
JW and GF acknowledge NASA grant GO3-14117. JW acknowledges support from NSFC 
grants 11443003 and 11473021.


\bibliography{ulx}

\end{document}